\documentclass[12pt,flexeq]{article}
\usepackage[cp1250]{inputenc}
\usepackage{amsmath, amsthm}
\usepackage{amsfonts}
\usepackage{amssymb}
\usepackage{verbatim}
\usepackage{xypic}
\usepackage{subcaption}
\usepackage{color}
\usepackage{todonotes}
\newcommand{\cmt}[1]{\todo[color=green!15!white, linecolor=black]{#1}}
\newcommand{\cmti}[1]{\todo[color=green!15!white,linecolor=black,inline]{#1}}
\newcommand{\tred}[1]{\textcolor{red}{#1}}
\makeatother
\newcommand{\ddiv}{\mathop {\rm div}\nolimits }
\newcommand{\curl}{\mathop {\rm curl}\nolimits }
\newcommand{\grad}{\mathop {\rm grad}\nolimits }
\newcommand{\TEM}{T^{\scriptscriptstyle EM}}
\newcommand{\QYK}{\Theta}
\newcommand{\rd}{{\rm d}}
\newcommand{\TBR}{T^{\scriptscriptstyle BR}}
\newcommand{\QBR}{CQ^{\scriptscriptstyle BR}}
\newcommand{\kolo}[1]{\vphantom{#1}\stackrel{\circ}{#1}\!\vphantom{#1}}
\newcommand{\dtwo}{\mathbf{\Delta}}

\newtheorem{sthm}{Proposition}[section]
\newtheorem{prf}{Proof}
\newtheorem{thm}{Theorem}[section]
\title{\bf Simple description of generalized electromagnetic and gravitational hopfions}
\author{Tomasz Smołka\thanks{E--mail: \texttt{Tomasz.Smolka@fuw.edu.pl}}\; and Jacek Jezierski\thanks{E--mail: \texttt{Jacek.Jezierski@fuw.edu.pl}} \\
	Department of Mathematical Methods in
	Physics, \\ University of Warsaw,
	ul. Pasteura 5, 02-093 Warszawa, Poland}
\begin{document}
	\numberwithin{equation}{section}
	\maketitle
\begin{abstract}
 The generalization of electromagnetic and gravitational hopfions is performed in terms of a complex scalar field. New definition of topological charge for linearized gravity is given. Quasi-local (super-)energy densities are compared for gravitational hopfion.
\end{abstract}
\begin{comment}
Generalized electromagnetic and gravitational hopfions are analyzed in terms of a complex scalar field. New definition of topological charge for spin-2 field is given.
\end{comment}
	\section{Introduction}
		A hopfion or Hopf soliton is a `solitonary' solution of spin-N field which has non-trivial topological properties. Field lines of the Hopf solution are related to the structure of Hopf fibration.  The characteristic structure of hopfion can be easily seen on the integration curves of the vector field (see \cite{irvine2008linked}). Briefly, the field lines are linked and form a knot. It cannot be unknoted without cutting.
		
		In 1977 Trautman \cite{trautman1977solutions} proposed the first electromagnetic solutions which were derived from the Hopf fibration. Ra\~nada developed them to propagating solutions in refs \cite{ranada1989topological,ranada1990knotted}. Last years, these little known solutions become more interesting because hopfions have successful applications in many areas of physics including electromagnetism \cite{ranada2004topological,irvine2008linked}, magnetohydrodynamics \cite{kamchatnov1982topological}, hadronic physics \cite{skyrme1962unified} and Bose--Einstein condensate \cite{kawaguchi2008knots}. The definition of hopfion was extended in \cite{thompson2015classification}. It includes a class of spin N-fields and uses this to classify the electromagnetic and gravitational hopfions by algebraic type.
		
		The paper has two main parts: the first is related to electromagnetic generalization of hopfion, the second one presents the linearized gravity case.
		
		In the first part, we briefly present a description of electromagnetic hopfion-like solution with the help of a complex scalar field\footnote{The E-M field can be equivalently represented by the complex scalar field $\Phi=(E+\imath B) \cdot \mathbf{r}$, where $E$ -- electric vector field, $B$ -- magnetic vector field, $\mathbf{r}$ -- position vector field and $\imath^2=-1$. Cf. equation (\ref{eq:wave_eq}) and the appendix \ref{Sec:scal_rep_EM_field}.} $\Phi$. The description of E-M field in terms of $\Phi$ is presented in the appendix \ref{Sec:scal_rep_EM_field}. We demonstrate a simple parametrization of such class of generalized hopfions by scalar wave function $\Phi$. The reconstruction of electromagnetic solution from such function is presented. Next, we show the condition for conservation of topological charge -- electric (magnetic) helicity in time in terms of $\Phi$.
		
		In the second part, the description of linearized gravity hopfions in terms of the complex scalar field $\Psi$ is presented.  The reconstruction of a gravitational hopfion from such complex, scalar function is performed. We propose a new definition of a topological charge for spin-2 field in analogy to the electromagnetic case. Hamiltonian energy for linearized gravity is discussed. A few quasi-local (super-)energy densities are presented in terms of the complex scalar $\Psi$. We compare such quasi-local (super-)energy densities for gravitational hopfion.
		
		To clarify the exposition, a full explanation of a complex, scalar framework for electromagnetism/linearized gravity has been placed in the appendix.
		\subsection{Notation}
		For convenience we use index notation with Einstein summation convention.  Minkowski spacetime is the background with the metric $g=-\mathrm{d} t^2+ \delta_{ab} \mathrm{d} x^{a} \mathrm{d} x^{b}$. The three-dimensional spatial metric is denoted by $\delta_{ab}$. Small latin indices, except $t$ and $r$, run spatial coordinates on $\Sigma=\{t=\mathrm{const.}\}$ slice. The distinguished $t$ and $r$ are respectively time and radius; $\mathbf{r}$ is a three-dimensional position vector. Capital letters represent the axial coordinates on the unit sphere. `,' denotes the partial derivative $\partial$. `$|$' is a three-dimensional spatial covariant derivative on $\{t=\mathrm{const.}\}$ surface and `$||$' denotes a  two-dimensional covariant derivative on the sphere {of radius $r$}. The two-dimensional trace is denoted by $\stackrel{(2)}{X}=g^{CD} X_{CD}$ and the two-dimensional traceless part ${\vphantom{X}\stackrel{\circ}{X}\!\vphantom{X}}_{AB}=X_{AB}-\frac12 g_{AB} \stackrel{(2)}{X}$. We will denote by $T_{...(\mu\nu)...}$ the symmetric part and by $T_{...[\mu\nu]...}$ the antisymmetric part of tensor $T_{...\mu\nu...}$ with respect to indices $\mu$ and $\nu$ (analogous symbols will be used for more indices).

	\section{Generalized hopfions in electrodynamics \label{sec:E-M_chapter}}
			\subsection{Class of generalized hopfions \label{sec:Rec_for_EM_field}}
				Consider a class of complex functions on the Minkowski background which are harmonic:
\begin{equation}
					\Box \Phi=0 \label{eq:wave_eq}
				\end{equation}
%Each of the presented wave function $\Phi$ generates a hopfion-like solution.
where $\Box$ is the d'Alembert operator in Minkowski spacetime.\\
There exists a bijection between electromagnetic solutions and such complex scalar fields.
For a given Riemann-Silberstein vector $Z:=E+\imath B$, complex combination of electric vector field $E$ and magnetic vector field $B$, we simply define
 $\Phi:=Z \cdot \mathbf{r}$ i.e. $\Phi$ is the scalar product of Riemann--Silberstein vector and position vector (c.f. equation (\ref{eq:RS}) and comments nearby).
To check the inverse mapping we need to show the reconstruction of the full EM data $Z$ from a wave function $\Phi$.
				
%				A distinguished set of solutions\marginpar{\tiny tu ciagle jest zle, chodzi o to, ze kazde rozwiazenie rownania Maxwella redukuje sie do rownania 2.1, to nie ma nic wspolnego z hopfionami!} of the following wave equation for a complex scalar function				
%				where $\Phi=Z \cdot r$ is the scalar product of Riemann--Silberstein vector $Z=E+\imath B$ and position vector (c.f. equation (\ref{eq:RS}) and comments nearby). $\Box$ is the d'Alembert operator in Minkowski spacetime.

From now, we restrict ourselves to use $(t,\theta, \varphi,r)$ coordinates\footnote{$t$ and $r$ denote respectively time and radial coordinate. $\theta$ and $\varphi$ parametrizes the two-sphere.} with metric \\$\delta_{ab}\mathrm{d} x^{a} \mathrm{d} x^{b}=\mathrm{d}r^2+r^2\left(\mathrm{d}\theta^2+\sin^{2} \theta \mathrm{d} \varphi^{2}\right)$.

The procedure {presented below} describes how to recover Riemann--Silberstein vector field $Z$ from $\Phi$. We would like to stress that the presented procedure can be used for any smooth solution of (\ref{eq:wave_eq}).
				
				 The definition of $\Phi$ and some of the vacuum Maxwell equations (\ref{eq:cMaxwell1}) and (\ref{eq:cMaxwell2}) in terms of  scalar $\Phi$ (in index notation) take the form
				\begin{eqnarray}
				\partial_{r} \left(r \Phi\right) &=&-r^2 {Z^{A}}_{||A} \label{eqn:2DMaxwell1} \\
				\partial_{t}\Phi&=& \imath \varepsilon^{ A B} Z_{A||B} \label{eqn:2DMaxwell2} \\
    			\Phi &=& r Z^r
				\end{eqnarray}
	where $\varepsilon^{AB}$ is a Levi-Civita tensor\footnote{{It can be defined by the formula $\varepsilon_{\alpha \beta \gamma \delta}=r^2 \sin \theta \epsilon_{\alpha \beta \gamma \delta}$, where
					\begin{equation}
					\epsilon_{\alpha \beta \gamma \delta}=\left \{\begin{array}{l}
					+1 \quad \mbox{if $\alpha \beta \gamma \delta$ is an even permutation of $\{t,\theta,\varphi,r\}$}\\
					-1 \quad \mbox{if $\alpha \beta \gamma \delta$ is an odd permutation of $\{t,\theta, \varphi,r\}$}\\
					\hspace{0.25 cm}0 \hspace{0.2 cm} \mbox{ in any other case}
					\end{array} \right. \nonumber
					\end{equation}
For lower dimensional case, we have $\varepsilon_{a b c}=\varepsilon_{t a b c}$ and $\varepsilon_{AB}=\varepsilon_{r AB}$. \label{L-C_tensor}}} on a sphere $t=$const., $r=$const.
		Hence, quasi-locally the above formulae enable one to reconstruct $Z$.
	More precisely, according to Hodge--Kodaira theory applied to differential forms on a sphere\footnote{See appendix \ref{sec:operation_on_S2}. There are no harmonic one-forms on a two-sphere.}, $Z_A {\rm d} x^A$    %,an axial part of $Z$,
can be decomposed into a gradient and co-gradient of some functions $\alpha$ and $\beta$
				\begin{equation}
				Z_{A}=\alpha_{,A}+\varepsilon_A{^B}\beta_{,B} \label{eq:Hodge_decomposition}
				\end{equation}
				The equations (\ref{eqn:2DMaxwell1})-(\ref{eq:Hodge_decomposition}) allow to obtain $\mathbf{\Delta} \alpha$ and $\mathbf{\Delta} \beta$. The two-dimensional Laplace operator $\mathbf{\Delta}$ on the unit sphere can be quasi-locally inverted with the help of methods which  are presented in appendix \ref{sec:math_sup}. From now, we restrict ourselves to the function $\Phi$ which  is a $l$-pole like in the formula (\ref{eq:Phi_Hopf}).
			   For convenience, we define the time-radius part $\phi$ of $\Phi$:
				\begin{equation}
				\Phi=\phi(t,r) Y_{l}(\theta,\varphi) \label{eq:time-radius_EM}
				\end{equation}
				where $Y_{l}$ is the $l$-th spherical harmonics -- eigenfunction of the two-dimensional Laplace operator on the unit sphere, i.e.
				\begin{equation}
				\mathbf{\Delta} Y_{l}=-l (l+1) Y_{l}
				\end{equation}
				We can split $\alpha$ and $\beta$ into multipoles. We highlight that the multipole decomposition is convenient to use in the examined case but the reconstruction procedure does not require multipole splitting in general.  (\ref{eq:time-radius_EM}) suggests that only one $l$-pole will be non-vanishing in the expansion
				\begin{equation}
				Z^{A}= a(t,r) (Y_{l }){^{,A}} + b(t,r) \varepsilon^{A B} (Y_{l})_{,B} \label{eqn:Z_l_pole}
				\end{equation}
				Combining (\ref{eqn:2DMaxwell1}) with (\ref{eqn:Z_l_pole}), the direct formula for complex scalar function $a(t,r)$ is obtained:
				\begin{equation}
				a(t,r)=\frac{ \partial_{r}\left(r \phi(t,r) \right)}{l (l+1) } \label{eq:a_EM}
				\end{equation}
				Analogically, using (\ref{eqn:2DMaxwell2}) and (\ref{eqn:Z_l_pole}), we obtain the function $b(t,r)$:
				\begin{equation}
				b(t,r)=- \imath \frac{r}{l (l+1)  } \partial_{t} \phi(t,r) \label{eq:b_EM}
				\end{equation}
				We reconstruct the two-dimensional part of $Z$. The radial component of $Z$ is algebraically related with $\Phi=Z^{r} r$. We recover the full form of $Z$ in that way.

				In the context of hopfions, the interesting set of solutions of (\ref{eq:wave_eq}) is
				\begin{equation}
				\Phi_{H}=\frac{r^{l} Y_{l}}{\left[r^2-(t-\imath)^2\right]^{l+1}} \label{eq:Phi_Hopf}
				\end{equation}
				The dipole solution from (\ref{eq:Phi_Hopf}) is related to Hopfion solution from \cite{thompson2015classification},  so we call (\ref{eq:Phi_Hopf}) generalized hopfions. The properties of solutions {(\ref{eq:Phi_Hopf})} are discussed in the sequel at the end of section~\ref{sec:E-M_chapter}.

			\subsection{Chandrasekhar--Kendall vector potential}
				A vector potential is defined up to a gradient of some function by the formula $Z=\curl V$ -- cf. appendix \ref{Sec:scal_rep_EM_field}, equation (\ref{eq:cMaxwell3}). The field $Z$ for presented class of generalized hopfions (\ref{eq:Phi_Hopf}) has simple multipole structure. It leads to a similar form of $V$. We propose for $V$ following \textit{ansatz}:
				\begin{eqnarray}
				V^{r}&=& s (t,r) Y_{l} \\
				V^{A}&=&p(t,r) (Y_{l})^{,A}+q(t,r) \varepsilon^{AB}(Y_{l})_{ ,B}
				\end{eqnarray}
				The above formulae and the Maxwell equation (\ref{eq:cMaxwell3}) imply %following
				\begin{eqnarray}
				r \phi(t,r) &=&l (l+1)q(t,r) \label{eq:Z_rA_r}\\
				a(t,r)&=&\partial_{r}[ q(t,r)] \\
				b(t,r)&=&s(t,r)-\partial_{r}[p(t,r)] \label{eq:Z_rA_r2}
				\end{eqnarray}
				{For solutions (\ref{eq:Phi_Hopf})}, freedom of choice of $V$ enables one to construct vector potential in Chandrasekhar--Kendall (C--K) form\footnote{Chandrasekhar--Kendall potential is part of a family of fields known as force-free fields and is of broad importance in plasma physics and fluid dynamics. See \cite{irvine2008linked} and the citations within.}. C--K potential is an eigenvector of the curl operator
				\begin{equation}
					Z= \lambda (t,r) V \label{eq:eigenpotential}
				\end{equation}
				where $\lambda (t,r)$ is a complex, scalar function. It leads to an overdetermined system of equations
				\begin{eqnarray}
				\phi(t,r) &=& \lambda s(t,r) \label{eq:eigen_Z}\\
				a(t,r)&=&\lambda p(t,r) \\
				b(t,r)&=&\lambda q(t,r) \label{eq:eigen_b_q}
				\end{eqnarray}
				It turns out that the equations (\ref{eq:Z_rA_r}-\ref{eq:Z_rA_r2}) and (\ref{eq:eigen_Z}-\ref{eq:eigen_b_q}) {for solutions (\ref{eq:Phi_Hopf})} are self-consistent. {For (\ref{eq:Phi_Hopf}), we introduce the time-radius part (\ref{eq:time-radius_EM}) denoted by $\phi_{H}(t,r)$.} The solutions are the following functions
				\begin{eqnarray}
				s(t,r)&=& \frac{\imath \phi_{H}(t,r)^{2}}{\partial_{t} \phi_{H}(t,r)} \\
				p(t,r)&=&\frac{\imath \phi_{H}(t,r) \partial_{r}(r \phi_{H}(t,r))}{l (l+1) \partial_{t} \phi_{H}(t,r)} \\
				q(t,r)&=& \frac{r \phi_{H}(t,r)}{l (l+1) }
				\end{eqnarray}
				which represent eigenvector of (\ref{eq:eigenpotential}) with the following eigenvalue:
				\begin{equation}
				\lambda (t,r)= - \imath \partial_{t} \ln(\phi_{H}(t,r))
				\end{equation}
		\subsection{Conservation of topological charge in time}
For electric and magnetic field fulfilling constraints one can introduce vector potentials:
\[ E=\curl C \, , \quad B = \curl A \, , \quad V:=C+\imath A \, , \quad Z=\curl V \]
			See appendix \ref{Sec:scal_rep_EM_field} for details. Topological charge is related to a number of linkedness and knotness of the integral curves of the electric $E$ (magnetic $B$) vector field. For electromagnetic field, electric helicity
			\begin{equation}
			h_{E}=\int_{\Sigma} C \cdot E  \label{eq:def_hel_E}
			\end{equation}
			and magnetic helicity
			\begin{equation}
			h_{M}=\int_{\Sigma} A \cdot B \label{eq:def_hel_M}
			\end{equation}
			are quantities which allow one to measure changes of the topological charge (see \cite{berger1999introduction}). $C$ and $A$ are vector potentials for $E$ and $B$ respectively (see appendix \ref{Sec:scal_rep_EM_field} for details). $\Sigma$ means the whole spatial space on a slice $\{t = \mathrm{const.}\}$ and $\cdot$ is a scalar product. It is convenient to present the helicities in terms of Riemann-Silberstein vector field $Z$ and its vector potential $V$:
			\begin{eqnarray}
			h_{E}+h_{M}&=&\int_{\Sigma} \Re\left(Z \cdot \overline{V}\right) = \Re\int_{\Sigma} Z^{a}\overline{V}^b \delta_{ab} \, \mathrm{d}^{3} x \label{eq:h_complex_bar}\\
			h_{E}-h_{M}&=&\int_{\Sigma} \Re\left(Z \cdot V\right) =\Re \int_{\Sigma} Z^{a}V^{b} \delta_{ab} \, \mathrm{d}^{3} x \label{eq:h_complex}
			\end{eqnarray}
			where $\overline{V}$ is the complex conjugate of $V$ and $\Re$ denotes the real part. Using the scalar description of E-M fields (appendix \ref{Sec:scal_rep_EM_field}) we can express total helicity (\ref{eq:h_complex_bar}) in terms of $\Phi$:
			\begin{equation}
			h_{E}+h_{M}=\int_{\Sigma} \Re \left[ \imath \left(\Phi \mathbf{\Delta}^{-1} \partial_{t} \overline{\Phi}-\overline{\Phi}\mathbf{\Delta}^{-1} \partial_{t} \Phi \right) \right] \label{eq:h_Phi_2}
			\end{equation}
			where $\imath^{2}=-1$ and $\mathbf{\Delta}^{-1}$ is an inverse operator to the two-dimensional Laplace operator on the unit sphere (see appendix \ref{sec:math_sup}).

The equations (\ref{eq:h_Phi_2}) and (\ref{eq:wave_eq}) imply conservation law for total helicity:
\begin{eqnarray}
\partial_{t} \left(	h_{E}+h_{M}\right) &=& \lim\limits_{R \to \infty}\int_{B(0,R)} \Re \left[ \imath \left(\Phi \mathbf{\Delta}^{-1} \partial^2_{t} \overline{\Phi}-\overline{\Phi}\mathbf{\Delta}^{-1} \partial^2_{t} \Phi \right) \right] \nonumber \\
&=&\lim\limits_{R \to \infty} \int_{\partial B(0,R)} \Re  \left[\imath \left(\Phi\mathbf{\Delta}^{-1} r^2 \partial_{r} \overline{\Phi}-  r^2 \partial_{r} \Phi\mathbf{\Delta}^{-1} \overline{\Phi} \right)\right]
\end{eqnarray}
			where $B(0,R)=\{x \in \Sigma:||x||\leq R\}$.
			 We assume the E-M fields are localized,\footnote{By localized we mean compactly supported or with fall off sufficiently fast which enables one to neglect boundary terms.}
hence the boundary terms at infinity can be neglected. In general, for the quantity $h_{E}-h_{M}$ (\ref{eq:h_complex}) we have no time dependence. However, in terms of $\Phi$ we have			
			\begin{equation}
			h_{E}-h_{M}=-2  \int_{\Sigma} \Re \left( \imath \Phi \mathbf{\Delta}^{-1} \partial_{t} \Phi \right) \label{eq:h_Phi_1}
			\end{equation}
			and
			\begin{equation}
			\partial_{t} \left(	h_{E}-h_{M}\right)= -2  \int_{\Sigma} \Re \partial_{t} \left( \imath \Phi \mathbf{\Delta}^{-1} \partial_{t} \Phi \right) \label{eq:h_EM_final_1}
%			\partial_{t} \left(	h_{E}+h_{M}\right)&\equiv&0 \label{eq:h_EM_final_2}
			\end{equation}
			which lead to the following
			\begin{sthm}
				For localized fields, the helicities (\ref{eq:def_hel_E}) and (\ref{eq:def_hel_M}) are preserved in time if and only if
				\begin{equation}
				\int_{\Sigma} \Re \partial_{t} \left(\imath \Phi \mathbf{\Delta}^{-1} \partial_{t} \Phi \right)=0 \label{eq:topo_charge_condition}
				\end{equation}
				\label{sthm:E_M_topo_conservation}				
			\end{sthm}
		The following quasi-local equality
		\begin{equation}
		\int_{\partial B(0,R)} Z \cdot Z=\int_{\partial B(0,R)} \partial_{t} \left(\Phi \mathbf{\Delta}^{-1} \partial_{t} \Phi \right) \label{eq:Z_Z}
		\end{equation}
		gives equivalence to the Ra\~nada result in \cite{ranada1989topological}.
		We highlight that $\cdot$ denotes scalar product without complex conjugate.
			\subsection{Discussion}
			The conservation of topological charge imposes an additional condition (\ref{eq:topo_charge_condition}) for solutions (\ref{eq:Phi_Hopf}). The integral in (\ref{eq:topo_charge_condition}) for solutions (\ref{eq:Phi_Hopf}) contains an integral of a square of a single multipole $Y_{l}$ over a two-dimensional sphere. $\int_{0}^{\pi} \mathrm{d} \theta \int_{0}^{2 \pi} \mathrm{d} \varphi (Y_{l})^2 $ is equal to zero for non-zero order $m$ of multipole. We denote $Y_{l}=Y_{lm}$ where $l$ and $m$ are respectively a degree and an order\footnote{Physicists usually use the naming convention which is associated with quantum mechanics. The degree of multipole is related to spin number and the order of multipole corresponds to magnetic spin number.} of multipole. Hence these values of $m$ for each $l$ lead to an E-M field which preserves the topological charge. Such E-M solution is a generalization of the null hopfion. For $l=1$ our solutions with the maximal {order} are equal (up to a constant) to the null hopfion described in \cite{thompson2015classification}. The case $l=1, m=0$ corresponds to non-null hopfion from \cite{thompson2015classification}.

	\section{Spin-2 field and generalized gravitational hopfions}
			Consider a weak gravitational field on the Minkowski background. The used complex scalar framework is related to the linearized Weyl tensor splitted into a tidal (gravito-electric) part $E_{kl}$ and frame-dragging
(gravito-magnetic) part $B_{kl}$ (see appendix \ref{sec:grav_E_B}). Both $E_{kl}$ and $B_{kl}$ are symmetric and traceless. With the help of the constraint equations,
			\begin{eqnarray}
				{E^{kl}}_{|l}&=&0 \\
				{B^{kl}}_{|l}&=&0
			\end{eqnarray}
			 we can quasi-locally describe the field in the terms of complex scalar field $\Psi$. See the appendix \ref{sec:Scalar_spin_2} for precise formulation and details. The used notation and denotings are presented in appendix \ref{sec:section_lin_grav}.
			 \subsection{Reconstruction for linearized gravity field}
			 The reconstruction for linearized gravity field is a generalization of the procedure for electromagnetic field described in the section \ref{sec:Rec_for_EM_field}. Constraint equations and the Hodge--Kodaira decomposition for two-dimensional tensors on a sphere (see appendices \ref{sec:spher_ident} and \ref{sec:Scalar_spin_2}) enable one to encode quasi-locally a spin-2 field into a complex scalar field. For a given $l$-pole field the reconstruction is simplified because the inverse operator to the two-dimensional Laplacian has a simple form. In the context of hopfions, we consider a class of complex scalar fields in the following form
			 \begin{equation}
			 \Psi_{H}= \frac{r^{l} Y_{l}}{\left[r^2-(t-\imath)^2\right]^{l+1}} \label{eq:Psi_hopf}
			 \end{equation}
			 for $l \geq 2$. For convenience, we define
			 \begin{equation}
				 \psi_{H}=\frac{r^{l}}{\left[r^2-(t-\imath)^2\right]^{l+1}}
			 \end{equation}
			 $\Psi_{H}=\psi_{H} Y_{l}$ is the same function as $\Phi_{H}$ (\ref{eq:Phi_Hopf}) for the set of generalized electromagnetic hopfions. For $l=2$, the solution (\ref{eq:Psi_hopf}) is related to gravitational hopfion\footnote{The type N hopfion from \cite{thompson2015classification} covers (up to a constant) with the solution from class (\ref{eq:Psi_hopf}) for $l=2$ and for the spherical harmonic with maximum spin number ($m=l=2$).}, so we call the set of solutions (\ref{eq:Phi_Hopf}) generalized gravitational hopfions.  $\Psi_{H}$ fulfills wave equation $\Box \Psi_{H}=0$ and represents gauge-invariant reduced data for %is equivalent to fulfillment
  linearized vacuum Einstein equation. For given $l$-pole field (\ref{eq:Psi_hopf}) the structure of reconstructed gravito-electromagnetic tensor $Z_{kl}$ is as follows:
			 \begin{eqnarray}
			 Z^{rr}&=&a_g(t,r) Y_{l} \\
			 Z^{rA}&=&b_{g}(t,r)(Y_{l})^{||A}+c_{g}(t,r) \varepsilon^{r AB}(Y_{l})_{||B} \\
			 \stackrel{(2)}{Z}&=&-a_g(t,r) Y_{l} \\
			 \stackrel{\circ}{Z}_{AB}&=&d_{g}(t,r) \left((Y_{l})_{||AB}-\frac{1}{2}\frac{g_{AB}}{r^{2}}\right)+e_{g}(t,r) (Y_{l})_{||C(A}{\varepsilon_{B)}}^{C}
			 \end{eqnarray}
			 where
			 \begin{eqnarray}
				 a_g(t,r)&=&\frac{\psi_{H}}{r^2} \\
				 b_{g}(t,r)&=&\frac{\partial_{r}(r \psi_{H})}{l (l+1) r}\\
				 c_{g}(t,r)&=&\frac{\partial_{t} \psi_{H}}{l (l+1)}\\
				 d_{g}(t,r)&=&\frac{\partial_{r}\left(r \partial_{r}\left(r \psi_{H}\right)\right)-\frac12 l(l+1) \psi_{H}}{l \left(l+1\right)\left[l\left(l+1\right) -2\right]}\\
				 e_{g}(t,r)&=&\frac{\partial_{r}\left(r^2 \partial_{t} \psi_{H} \right)}{l \left(l+1\right)\left[l\left(l+1\right) -2\right]}		
			 \end{eqnarray}
			 which are similar to (\ref{eq:a_EM})-(\ref{eq:b_EM}) for electromagnetic case.
\subsection{Hamiltonian energy for linearized gravity \label{sec:spin_2_energy}}

 %In our opinion
% the functional \eq{EMV}
% seems to be the most natural one
%because it contains only first (radial and time) derivative
%of our quasi-local variables. Moreover, one can easily check
%the conservation law for compactly supported data
%straightforward from the wave equation
%which is fulfilled by our quasi-local unconstrained degrees
%of freedom (see Appendix \ref{xyEH}). The functional $\QYK_0$
%is also very
%which are close to the energy functional proposed in
In \cite{jezierski1990localization} (see also \cite{jezierski1995relation} and \cite{JJschwarzl}) one of us proposed energy functional ${\cal H}$
which takes the following form in Minkowski spacetime:
\begin{eqnarray}
 {\cal H} & = & \nonumber
 \frac1{32\pi} \int_{\Sigma}
 \Bigl[ (r\dot{\bf x})
 \dtwo ^{-1}(\dtwo +2)^{-1} (r\dot{\bf x}) + (r\dot{\bf y})
 \dtwo ^{-1}(\dtwo +2)^{-1} (r\dot{\bf y})
\\ & &  \label{Hxy2} \hphantom{\frac1{4} \int_{\Sigma}}
 + (r{\bf x})_{,r}
 \dtwo ^{-1}(\dtwo +2)^{-1} (r{\bf x})_{,r} -
  {\bf x}  (\dtwo +2)^{-1}{\bf x}
\\ & &  \nonumber \hphantom{\frac1{4} \int_{\Sigma}} +  (r{\bf y})_{,r}
 \dtwo ^{-1}(\dtwo +2)^{-1} (r{\bf y})_{,r} - {\bf y}
 (\dtwo +2)^{-1} {\bf y}  \Bigr]
\rd r \sin\theta\rd\theta\rd\varphi %\, .
  \end{eqnarray}
  %\cmti{Eq. below update}
  or in terms of $\Psi= {\mathbf x} + \imath {\mathbf y}$
\begin{eqnarray}
{\cal H} & = &  \label{Hxy}
\frac1{32\pi} \int_{\Sigma}
\Bigl[ (r\partial_{t} \Psi)
\dtwo ^{-1}(\dtwo +2)^{-1} (r\partial_{t} \overline{\Psi})
\\ & & \nonumber  \hphantom{\frac1{4} \int_{\Sigma}}
+ (r\Psi)_{,r}
\dtwo ^{-1}(\dtwo +2)^{-1} (r \overline{\Psi})_{,r} -
\Psi  (\dtwo +2)^{-1} \overline{\Psi}   \Bigr]
\rd r \sin\theta\rd\theta\rd\varphi %\, .
\end{eqnarray}
%and $r^2 V^{(+)}=r^2 V^{(-)}= -\dtwo$.
The formula has its origins in the canonical (Hamiltonian) formulation of the linearized theory of gravity. In this sense it describes a true energy of linearized gravitational field. In section \ref{sec:energy_comparision} we remind another two super-energy functionals\footnote{We remind a quasi-local densities of such energy functionals. The diffrence is only to integrate over the radial coordinate i.e. $\Theta_{0}=\int_{0}^{\infty} U_{\Theta_{0}} \mathrm{d} r$} $\QYK_0$ (\ref{EMV}) (see also \cite{jezierski2002cyk}) and super-energy (\ref{eq:q_l_super_energy}) which arises for spin-2 field in a natural way.
In particular, the integrals (\ref{EMV}) and (\ref{Hxy}) differ by the operator
$(\dtwo +2)^{-1}$, hence for each spherical mode
(i.e. after spherical harmonics decomposition) they
 are proportional to each other. Hamiltonians for whom functions in multipole expansions differ by a constant multiplicative factor lead to the same dynamics. We will discuss in a separate paper \cite{jjjkenergy} %(it is not obvious)
 how the functional ${\cal H}$ is related to the following expression:
 \begin{eqnarray} \nonumber
			 16\pi \overline{\cal H} &=&\int_{\Sigma}
\left( E^{a b} (-\triangle)^{-1} E_{ab} + B^{a b} (-\triangle)^{-1} B_{ab}\right) \\
 &=& \iint_{\Sigma \times \Sigma} \left[
 \frac{ E^{a b}\left(\mathbf{r'}\right) E_{ab}\left(\mathbf{r''}\right)}{4\pi\|\mathbf{r'}-\mathbf{r''}\|}
 + \frac{ B^{a b}\left(\mathbf{r'}\right) B_{ab}\left(\mathbf{r''}\right)}{4\pi\|\mathbf{r'}-\mathbf{r''}\|}
 \right]
 \mathrm{d}\mathbf{r'} \mathrm{d}\mathbf{r''} \label{eq:Hxy2} \\
  %\end{eqnarray}
  %\cmti{Eq. below update}
  %\begin{eqnarray} \nonumber
  %16\pi \overline{\cal H} &=&\int_{\Sigma}
  %\left( E^{a b} (-\triangle)^{-1} E_{ab} + B^{a b} (-\triangle)^{-1} B_{ab}\right) \\
  &=&\int_{\Sigma}
  \left( Z^{a b} (-\triangle)^{-1} \overline{Z}_{ab}\right) \nonumber \\
  &=& \iint_{\Sigma \times \Sigma} \left[
  \frac{ Z^{a b}\left(\mathbf{r'}\right) \overline{Z}_{ab}\left(\mathbf{r''}\right)}{4\pi\|\mathbf{r'}-\mathbf{r''}\|}
  \right]
  \mathrm{d}\mathbf{r'} \mathrm{d}\mathbf{r''} \label{eq:Hxy}
  \end{eqnarray}
  which is proposed by I. Bialynicki-Birula \cite{bialynicki2015quantum} and has a nice property -- it is manifestly covariant with respect to the Euclidean group. In the future we also plan to incorporate boundary terms because we want to generalize the above formulae to finite region with boundary.

%\subsection{Energy in terms of Weyl tensor ala IBB}
Let us consider localized initial data on $\Sigma$, i.e. compactly supported or with fall off sufficiently fast which enables one to neglect boundary terms.
The following theorem (to be presented in detail in \cite{jjjkenergy})
\begin{thm}
For localized data ${\cal H}= \overline{\cal H}$.
\end{thm}
\noindent can be checked as follows:
\begin{proof}
{Let us observe that} ${\bf x}= 2 x^k x^l E_{kl}$, ${\bf y}= 2 x^k x^l B_{kl}$.
If we introduce transverse--traceless potentials\footnote{Transverse--traceless  symmetric tensor-field $h_{kl}$ means $h_{kl}\delta^{kl}=0$ and ${h^{kl}}_{|l}=0$.} ${\tt e}$ and ${\tt h}$:
\[ -\triangle {\tt e}_{kl} = E_{kl}\, , \quad -\triangle {\tt h}_{kl} = B_{kl} \,  \]
where $\triangle$ is the three-dimensional Laplacian\footnote{In Cartesian coordinates it is simply $\displaystyle \triangle = \sum_{i=1}^3 \left(\frac{\partial}{\partial x^i}\right)^2$.}, then for
$ {\bf a}:= 2 x^k x^l {\tt e}_{kl}$, $ {\bf b}:= 2 x^k x^l {\tt h}_{kl}$
we get
\[ -\triangle {\bf a} = {\bf x} \, , \quad -\triangle {\bf b} = {\bf y} \,  \]
Moreover, for finite region $V\subset\Sigma$
\begin{eqnarray}
 16\pi \overline{\cal H}_V & := &
\int_V   \bigl({\tt e}_{kl}E^{kl} + {\tt h}_{kl}B^{kl}\bigr)\rd^3 x
%r^2\sin\theta\rd r \rd\theta\rd\varphi
 \\ &= & \nonumber \frac12 \int_V \frac1{r^2}\Bigl(
 (r\dot{\bf a})(-\dtwo)^{-1}(r\dot{\bf x})
 + \partial_r(r{\bf a})(-\dtwo)^{-1}\partial_r(r{\bf x}) + \frac12{\bf a}{\bf x}
 \\ & & \nonumber
 + (r\dot{\bf b})(-\dtwo)^{-1}(r\dot{\bf y})
 + \partial_r(r{\bf b})(-\dtwo)^{-1}\partial_r(r{\bf y}) + \frac12{\bf b}{\bf y}
 \Bigr)\rd r \sin\theta\rd\theta\rd\varphi  \\ & & \label{enIBB}
 + \frac12 \int_V \frac1{r^2}\Bigl[
   \partial_r(r^2\dot{\bf a})\dtwo^{-1}(\dtwo +2)^{-1}
   \partial_r(r^2\dot{\bf x}) + \frac14 {\bf a}{\bf x}  \\ & & \nonumber
   + \bigl(
   \partial_r [ r\partial_r(r{\bf a})] +\frac12\dtwo{\bf a}\bigr)
   \dtwo^{-1}(\dtwo +2)^{-1} \bigl(
   \partial_r [ r\partial_r(r{\bf x})] +\frac12\dtwo{\bf x}\bigr)
    \\ & & \nonumber
   + \bigl(
   \partial_r [ r\partial_r(r{\bf b})] +\frac12\dtwo{\bf b}\bigr)
   \dtwo^{-1}(\dtwo +2)^{-1} \bigl(
   \partial_r [ r\partial_r(r{\bf y})] +\frac12\dtwo{\bf y}\bigr)
   \\ & & \nonumber +
   \partial_r(r^2\dot{\bf b})\dtwo^{-1}(\dtwo +2)^{-1}
   \partial_r(r^2\dot{\bf y}) + \frac14 {\bf b}{\bf y}
   \Bigr]\rd r \sin\theta\rd\theta\rd\varphi
\end{eqnarray}
Now, we have to integrate by parts many times and finally we obtain energy (\ref{Hxy}) up to boundary terms
\begin{eqnarray}
 16\pi \overline{\cal H}_V & = &
%\int_V   \bigl({\tt e}_{kl}E^{kl} + {\tt h}_{kl}H^{kl}\bigr)\rd^3 x
%r^2\sin\theta\rd r \rd\theta\rd\varphi
  \nonumber \frac12 \int_V \Bigl[
 (-r^2\triangle \dot{\bf a})\dtwo^{-1}(\dtwo +2)^{-1}\dot{\bf x}
 + \partial_r(-r\triangle {\bf a})\dtwo^{-1}(\dtwo +2)^{-1}\partial_r(r{\bf x})
  \\ & & + \triangle{\bf a}(\dtwo +2)^{-1}{\bf x}
 \nonumber  %\hspace*{-1cm}
 + \partial_r(-r\triangle {\bf b})\dtwo^{-1}(\dtwo +2)^{-1}\partial_r(r{\bf y}) \\ \nonumber
  & & + (-r^2\triangle \dot{\bf b})\dtwo^{-1}(\dtwo +2)^{-1}\dot{\bf y} + \triangle{\bf b}(\dtwo +2)^{-1}{\bf y}
 \Bigr]\rd r \sin\theta\rd\theta\rd\varphi
 \\ & & \label{enIBB2}
 + \frac12 \int_{\partial V} \Bigl[
   \partial_r(r^2\dot{\bf a})\dtwo^{-1}(\dtwo +2)^{-1}\dot{\bf x}
   -\frac1{2r^2} \partial_r(r^2{\bf a})(\dtwo +2)^{-1}{\bf x}\\ & & \nonumber
   + \Bigl(r\triangle{\bf a}+\frac1{r}\partial_r(r{\bf a}) - \frac1{2r}\dtwo{\bf a} \Bigr)
   \dtwo^{-1}(\dtwo +2)^{-1} \partial_r(r{\bf x})
    \\ & & \nonumber
   + \partial_r(r^2\dot{\bf b})\dtwo^{-1}(\dtwo +2)^{-1}\dot{\bf y}
   -\frac1{2r^2} \partial_r(r^2{\bf b})(\dtwo +2)^{-1}{\bf y}\\ & & \nonumber
   + \Bigl(r\triangle{\bf b}+\frac1{r}\partial_r(r{\bf b}) - \frac1{2r}\dtwo{\bf b} \Bigr)
   \dtwo^{-1}(\dtwo +2)^{-1} \partial_r(r{\bf y})
   \Bigr] \sin\theta\rd\theta\rd\varphi %\, .
\end{eqnarray}
More precisely, the volume term in the above formula equals $16\pi {\cal H}$ given by (\ref{Hxy}).
\end{proof}
	\subsection{Quasi-local (super-)energy density for spin-2 field}
		We present quasi-local (q-l)  energy and super-energy densities for spin-2 field and linearized gravity. By q-l density we mean a functional which is an integral over a two-dimensional topological sphere. In the paper, we calculate q-l densities over $\{t=\mathrm{const.}, r=\mathrm{const.}\}$ surface. A few of analyzed energies (for example $\mathcal{H}$ (\ref{Hxy})) are defined with the help of q-l integral operator. They do not have density which can be calculated locally at point. The q-l (super-)energy densities listed below are presented in general form -- they are valid for every localized weak gravitational field represented as a complex harmonic function $\Psi$. The compared q-l (super-)energy densities can be organized as follows:
		\begin{enumerate}
			\item Related to the canonical (Hamiltonian) theory:
			\begin{enumerate}
				\item The q-l energy density of hamiltonian energy $\mathcal{H}$ (\ref{Hxy}) derived from the canonical formulation of linearized theory of gravity.
				\begin{eqnarray}
				U_{\mathcal{H}}&=&\label{eq:den_H} \frac1{32\pi} \int_{S(t,r)} \sin \theta \Bigl[ (r\partial_{t} \Psi)
				\dtwo ^{-1}(\dtwo +2)^{-1} (r\partial_{t} \overline{\Psi})
				\\ & &   \hphantom{\frac1{4} \int_{\Sigma}}
				+ (r\Psi)_{,r}
				\dtwo ^{-1}(\dtwo +2)^{-1} (r \overline{\Psi})_{,r} -
				\Psi  (\dtwo +2)^{-1} \overline{\Psi}   \Bigr]   \nonumber
				\end{eqnarray}
				where $S(t,r)$ denotes $\{t=\mathrm{const.}, r=\mathrm{const.}\}$ surface.
				\item $\Theta_{0}$ functional is obtained with the help of Conformal Yano--Killing (CYK) tensors. The contraction of CYK tensor $Q^{\mu \nu}$ with Weyl tensor $W_{\mu \nu \alpha \beta}$ is a two-form $F_{\alpha \beta}^{(Q)}=Q^{\mu \nu}W_{\mu \nu \alpha \beta}$, where $Q^{\mu \nu} \partial_{x^{\mu}} \wedge \partial_{x^{\nu}}=\mathcal{D} \wedge \partial_{t}$ is a CYK tensor for Minkowski spacetime and $\mathcal{D}=x^{\nu} \partial_{\nu}$ is a generator of dilatations in Minkowski spacetime. $F^{(Q)}_{\alpha \beta}$ fulfills vacuum Maxwell equations.  $\Theta_{0}$ is an electromagnetic energy calculated for $F^{(Q)}_{\alpha \beta}$ from stress-energy tensor
				\begin{equation}
				T^{\rm\scriptscriptstyle EM}_{\mu\nu}(F) := \frac12 \left( F_{\mu\sigma}F_\nu{^\sigma}
				+ F^*{_{\mu\sigma}}F^*{_\nu{^\sigma}}
				\right)
				\end{equation}
				where $F{^*}^{\mu\lambda}=\frac12\varepsilon^{\mu\lambda\rho\sigma}
				F_{\rho\sigma}$.
				See \cite{jezierski2002cyk} for details. The q-l density of $\Theta_{0}$ is
				\begin{eqnarray}
				4 \pi U_{\Theta_{0}}&=&\int_{S(t,r)} \TEM\bigl(\partial_{t},\partial_{t},
				F(W,{\cal D}\wedge{\cal T}_t)\bigr)r^2\sin\theta
				\rd\theta\rd\varphi \nonumber \\
				&=& \frac12 \int_{S(t,r)} r^2 \left(
				E_{kr}E^{kr} + B_{kr}B^{kr} \right)r^2\sin\theta
				\rd\theta\rd\varphi \nonumber \\
				& =&  \frac14 \int_{S(t,r)} \left[
				\partial_t(r{\Psi})(-\mathbf{\Delta})^{-1}\partial_t(r \overline{\Psi}) \right. \nonumber \\
				& &\left. + \partial_r(r \Psi)(-\mathbf{\Delta})^{-1}\partial_r(r \overline{\Psi}) + \Psi \overline{\Psi}
				\right] \sin \theta \rd\theta\rd\varphi \label{EMV}
				\end{eqnarray}
				\item We compare q-l energy densities for linearized gravity with q-l electromagnetic energy densities for the corresponding electromagnetic solution (compare (\ref{eq:Psi_solut_quadru}) and (\ref{eq:Phi_sol_quadru}) ). Let us define
				\begin{eqnarray}
				F_{1}(\Phi)&:=&\Bigl[ (r\partial_{t} \Phi) \nonumber
				(-\dtwo ^{-1})(r\partial_{t} \overline{\Phi})
				\\ & &  %\hphantom{\frac1{4} \int_{\Sigma}}
				+ (r\Phi)_{,r}
				(-\dtwo ^{-1}) (r \overline{\Phi})_{,r} +
				\Phi \overline{\Phi}   \Bigr]
				\sin\theta \\
				F_{2}(\Phi)&:=& \left[ \partial_{r}(r \Phi) \mathbf{\Delta}^{-1} r \partial_{t} \overline{\Phi}+ \partial_{r}(r \overline{\Phi}) \mathbf{\Delta}^{-1} r \partial_{t} \Phi \right] \sin \theta
				\end{eqnarray}							
				The electromagnetic q-l energy density in terms of electromagnetic scalar $\Phi$ is equal to
				\begin{eqnarray}
				4 \pi U_{EM}&=& \int_{S(t,r)} \TEM\bigl(\partial_{t},\partial_{t},
				\Phi\bigr)r^2\sin\theta
				\rd\theta\rd\varphi \nonumber \\
				&=& \frac14 \int_{S(t,r)} F_{1}(\Phi) \rd \theta \rd \varphi \label{eq:EM_ener_den} 			 
				\end{eqnarray}
				The electromagnetic q-l energy density for the conformal field
				\begin{equation}
				\mathcal{K}=2 r t \partial_{r} +\left(t^2+r^2 \right) \partial_{t} \label{eq:conf_field_K}
				\end{equation}
				is the following
				\begin{eqnarray}
				4 \pi U_{CEM}&=&\int_{S(t,r)} T^{EM}(\mathcal{K},\partial_{t}, \Phi) r^2\sin\theta
				\rd\theta\rd\varphi \nonumber \\
				&=& \frac{1}{4} \int_{S(t,r)}  \left[ \left(r^2 +t^2 \right) F_{1}(\Phi) +2 r t F_{2}(\Phi) \right]
				\rd\theta\rd\varphi \label{eq:C_EM_ener_den}
				\end{eqnarray}						
			\end{enumerate}
			\item Associated to Bel--Robinson tensor. The Bel--Robinson tensor has the structure
			\begin{equation}
			T^{\scriptscriptstyle BR}_{\mu\nu\kappa\lambda} :=
			W_{\mu\rho\kappa\sigma} W_\nu{^\rho}{_\lambda}{^\sigma} +
			{W^*}_{\mu\rho\kappa\sigma} {W^*}_\nu{^\rho}{_\lambda}{^\sigma} \label{eq:B_L}
			\end{equation}
			where $(W^*)_{\alpha\beta\gamma\delta}=\frac 12 W_{\alpha\beta}{^{\mu\nu}}
			\varepsilon_{\mu \nu \gamma\delta}$. The spin-2 field equations
(\ref{eq:Z_costrain}) and (\ref{eq:Z_dynamical}) remain invariant under the global $U(1)$ transformation $Z^{kl} \to e^{i \alpha}Z^{kl}$.
%It shows the existence of the spatial duality of the
%level of physically relevant gravito-electromagnetic field in vacuum case.
%The super-energy density $u_{S}$ (\ref{eq:super_energ_den_u}) is a charge
%related to the dual symmetry.
 The duality invariance\footnote{Introducing $\mathcal{W}_{\alpha \beta \gamma \delta}=W_{\alpha \beta \gamma \delta}+\imath {^*} W_{\alpha \beta \gamma \delta}$, Bell-Robinson tensor (\ref{eq:B_L}) has the form $T^{\scriptscriptstyle BR}_{\mu\nu\kappa\lambda} :=
 	\mathcal{W}_{\mu\rho\kappa\sigma} \overline{\mathcal{W}}_\nu{^\rho}{_\lambda}{^\sigma}$. All components of $\mathcal{W}_{\mu\rho\kappa\sigma}$ depend linearly on $Z_{kl}$, there are no anti-linear complex conjugate terms $\overline{Z}_{kl}$. Hence all components of Bell-Robinson tensor are proportional to ``$Z \overline{Z}$'' which are invariant under $Z^{kl} \to e^{i \alpha}Z^{kl}$ transformation.} is a property of Bel--Robinson tensor. The q-l density of super-energy fulfills
			\begin{eqnarray}
			4 \pi U_{S}&=&\int_{S(t,r)} \frac{1}{2} T^{BR}(\partial_{t},\partial_{t},\partial_{t},\partial_{t},\Psi) r^2 \sin \theta \rd\theta\rd\varphi \nonumber \\
			&=&\int_{S(t,r)} u_{S} r^2 \sin \theta \rd\theta\rd\varphi \nonumber \\
			&=&\frac{1}{4}\int_{S(t,r)} F_{3}(\Psi) \label{eq:q_l_super_energy}
			\end{eqnarray}
			where $F_{3}(\Psi)$ is given by (\ref{eq:den_S}). Let us introduce
			\begin{eqnarray}
			F_{3}(\Psi)&:=& \nonumber \frac{1}{r^2}\left\{
			(r\partial_{t} \Psi)(-\mathbf{\Delta})^{-1}(r\partial_{t} \overline{\Psi})
			+ \partial_r(r \Psi)(-\mathbf{\Delta})^{-1}\partial_r(r \overline{\Psi}) + \frac12 \Psi \overline{\Psi} \right. \\
			& & \nonumber
			+\bigl(
			\partial_r [ r\partial_r(r\Psi)] +\frac12\mathbf{\Delta}\Psi\bigr)
			\mathbf{\Delta}^{-1}(\mathbf{\Delta} +2)^{-1} \bigl(
			\partial_r [ r\partial_r(r\overline{\Psi})] +\frac12\mathbf{\Delta}\overline{\Psi}\bigr)
			\\ & &\left.
			+\partial_r(r^2 \partial_{t} \Psi)\mathbf{\Delta}^{-1}(\mathbf{\Delta} +2)^{-1}
			\partial_r(r^2 \partial_{t} \overline{\Psi}) + \frac14 \Psi \overline{\Psi} \right\} \sin \theta \label{eq:den_S} \\
			F_{4}(\Psi)&:=&\left\{ \frac{1}{2} \left[\partial_r(r \Psi)(-\mathbf{\Delta})^{-1}(r\partial_{t} \overline{\Psi})
			+ \partial_r(r \overline{\Psi})(-\mathbf{\Delta})^{-1}(r\partial_{t} \Psi) \right] \right. \\
			& & \nonumber
			+\bigl(
			\partial_r [ r\partial_r(r\Psi)] +\frac12\mathbf{\Delta}\Psi\bigr)
			\mathbf{\Delta}^{-1}(\mathbf{\Delta} +2)^{-1} \partial_r(r^2 \partial_{t} \overline{\Psi})
			\\ & &\left.
			+ \bigl(
			\partial_r [ r\partial_r(r\overline{\Psi})] \bigr)\mathbf{\Delta}^{-1}(\mathbf{\Delta} +2)^{-1}
			\partial_r(r^2 \partial_{t} \Psi) \right\} \sin \theta
			\end{eqnarray}
			The Bel--Robinson charge for a conformal field is as follows
			\begin{eqnarray}
			4 \pi U_{CS}&=&\int_{S(t,r)} \frac{1}{2}T^{BR}(\mathcal{K},\partial_{t},\partial_{t},\partial_{t},\Psi) r^2 \sin \theta \rd\theta\rd\varphi \nonumber \\
			&=&\frac{1}{4} \int_{S(t,r)} \left[ \left(t^2+r^2 \right) F_{3}(\Psi)+ 2 r t F_{4}(\Psi) \right] \rd\theta\rd\varphi \label{eq:q_l_c_sup_ener}
			\end{eqnarray}
			where conformal field $\mathcal{K}$ is defined by (\ref{eq:conf_field_K}).
		\end{enumerate}
		\subsection{Comparison of the energies for hopfions \label{sec:energy_comparision}}
			In \cite{thompson2015classification}, the following super-energy density
			\begin{equation}
				u_{S}=\frac{E_{ab} E^{ab}+B_{ab}B^{ab}}{2} \label{eq:super_energ_den_u}
			\end{equation}
			has been calculated for gravitational type N hopfion. We highlight that type N hopfion overlap (up to a constant) with the solution from our class (\ref{eq:Psi_hopf}) for $l=2$ and for the spherical harmonic with maximal order ($m=l=2$). For such quadrupole solution
			\begin{equation}
			\Psi_{q} := \frac{r^{2} Y_{22}}{\left[r^2-(t-\imath)^2\right]^{3}} \label{eq:Psi_solut_quadru}
			\end{equation}
			we analyze q-l (super-)energy densities for linearized gravity (\ref{eq:den_H}), (\ref{EMV}), (\ref{eq:q_l_super_energy}) and (\ref{eq:q_l_c_sup_ener}) which are presented in the previous section. We compare them with the electromagnetic q-l energy densities (\ref{eq:EM_ener_den}) and (\ref{eq:C_EM_ener_den}) for the corresponding to $\Psi_{q}$ (\ref{eq:Psi_solut_quadru}) electromagnetic quadrupole solution
			\begin{equation}
			\Phi_{q}:=\frac{r^{2} Y_{22}}{\left[r^2-(t-\imath)^2\right]^{3}} \label{eq:Phi_sol_quadru}
			\end{equation}
			Let us define
			\begin{eqnarray}
			\xi(t,r)&:=&\frac{r^4}{\left((r+t)^2+1\right)^4 \left((r-t)^2+1\right)^4}\left[t^4+(\frac{14}{5}r^2+2)t^2+(r^2+1)^2\right] \label{eq:xi} \\
			\kappa(t,r) &:=&\frac{r^5 t}{\left((r+t)^2+1\right)^4 \left((r-t)^2+1\right)^4}\left[r^2+t^2+1 \right]  \\
			\eta(t,r)&:=&\frac{r^2}{\left((r+t)^2+1\right)^5 \left((r-t)^2+1\right)^5}\left[r^8+(12 t^2+4) r^6\right. \label{eq:eta} \\
			& &\left. +\left(\frac{126}{5} t^4+28 t^2+6\right)r^4 +(12 t^6+28 t^4+20t^2+4)r^2+(t^2+1)^4 \right]  \nonumber\\			
			\tau(t,r) &:=&\frac{t r^3 \left(r^2+t^2+1 \right)}{\left((r+t)^2+1\right)^5 \left((r-t)^2+1\right)^5}  \left[t^4+\left(\frac{22}{5} r^2+2\right) t^2+(r^2+1)^2 \right]
			\end{eqnarray}
			The results for quadrupole hopfion are the following:
			\begin{eqnarray}
				U_{\mathcal{H}}(\Psi_{q})&=&\frac{1}{24} \xi(t,r) \\ \label{eq:q-l_den_hamilonian}
				U_{\Theta_{0}}(\Psi_{q})&=&\frac{1}{3}\xi(t,r) \\
				U_{EM}(\Phi_{q})&=&\frac{1}{3}\xi(t,r)\\ \label{eq:q-l_den_E-M_energy}
				U_{S}(\Psi_{q})&=&\frac{1}{2} \eta(t,r)
				\end{eqnarray}
				\begin{eqnarray}
				U_{CEM}(\Phi_{q})&=&\frac{4}{15} \left[\frac{5}{4} \left(t^2+r^2\right) \xi(t,r)-6 r t \kappa(t,r) \right] \\
					U_{CS}(\Psi_{q})&=&\frac{1}{2}\left[ \left(t^2+r^2\right) \eta(t,r)+2 r t \tau (t,r) \right]
			\end{eqnarray}
			\begin{comment}
			 we compare the following energy and super-energy functionals: ${\cal H}$ defined by (\ref{Hxy}), $\QYK_0$--(\ref{EMV}),  $\QBR_0$--(\ref{BRV}), $u_{s}$--(\ref{eq:super_energ_den_u}).  We confront them with the energy of the quadrupole $(l=2)$ solution (\ref{eq:Phi_Hopf}) for electromagnetism. We can not compare energies and super-energy densities directly because they are not the same type of objects\footnote{${\cal H}$ and $\QYK_0$ are quasi-local quantities -- they can be obtained only on a topological two-sphere as a Gaussian integral. We can define a `quasi-local density' for a foliation of spheres in that way.}. We consider quasi-local densities on spheres $\{t=\mathrm{const.},r=\mathrm{const.} \}$ for the above mentioned energies and super-energies. \tred{Roughly, we integrate only on those surfaces.} \cmt{Be more precise} \\
			 The results and technical details are presented in appendix \ref{sec:En_com}. The quasi-local energy densities, denoted by $U$ with index, are  given by (\ref{eq:q-l_den_hamilonian})-(\ref{eq:q-l_den_E-M_energy}).
			 \end{comment}
			  One can observe the following:
			 \begin{enumerate}
			 	\item The q-l energy densities can be divided into two sets:
			 	\begin{eqnarray}
			 		X_{1}&=&\{U_{\mathcal{H}}, U_{\Theta_{0}}, U_{EM}; U_{S}\} \\
			 		X_{2}&=&\{U_{CEM}, U_{CS}\}
			 	\end{eqnarray} Functions in each set have similar properties. It means:
			 	\begin{enumerate}
			 		\item In the set $X_{1}$ we can distinguish a subset $\{U_{\mathcal{H}},U_{\Theta_{0}},U_{EM}\}$. Q-l (super-) energy densities in the subset differ by a multiplicative constant. Simple, single-multipole structure of the solutions (\ref{eq:Psi_solut_quadru}) and (\ref{eq:Phi_sol_quadru}) is responsible for proportionality of q-l (super-)energy densities in the subset. For solutions with the richer multipole structure relations between the densities will be more complicated.
			 		\item The set $X_{2}$ contains q-l energy densities for the conformal field $\mathcal{K}$ (\ref{eq:conf_field_K}). For $t=0$, the conformal q-l densities are proportional to theirs counterparts for $\partial_{t}$ field. $r^2$ is the proportional factor
			 		\begin{eqnarray*}
			 		U_{CEM}(\Phi_{q},t=0)&=&r^2 U_{EM} (\Phi_{q},t=0) \\
			 		U_{CS}(\Psi_{q},t=0)&=&r^2 U_{EM}(\Psi_{q},t=0)
			 		\end{eqnarray*}
			 	\end{enumerate}
		 	\item All the above presented q-l (super-)energy densities are localized on light cones for large $t$ and $r$.
			 \end{enumerate}
		
		\subsection{Topological charge}
			We were not able to find a definition of a topological charge for weak gravitational field in the literature. We propose a quantity which can be a good candidate for a topological charge and investigate its properties. Consider the following non-local objects:
			\begin{eqnarray}
			h_{GE}&=&\int_{\Sigma} E^{a b} (-\triangle^{-1}) S_{ab}=\iint_{\Sigma \times \Sigma}\frac{ E^{a b}\left(\mathbf{r'}\right) S_{ab}\left(\mathbf{r''}\right)}{4\pi\|\mathbf{r'}-\mathbf{r''}\|} \mathrm{d}\mathbf{r'} \mathrm{d}\mathbf{r''} \label{eq:def_H_E} \\
			h_{GB}&=&\int_{\Sigma} B^{a b} (-\triangle^{-1}) P_{ab}=\iint_{\Sigma \times \Sigma}\frac{ B^{a b}\left(\mathbf{r'}\right) P_{ab}\left(\mathbf{r''}\right)}{4\pi\|\mathbf{r'}-\mathbf{r''}\|} \mathrm{d}\mathbf{r'} \mathrm{d}\mathbf{r''}\label{eq:def_H_B}
			\end{eqnarray}
			where $\triangle^{-1}$ is an inverse operator to the three-dimensional Laplacian $\triangle$ (details in appendix \ref{sec:math_sup}). {$P_{ab}$ and $S_{ab}$ are respectively ADM momentum and its dual counterpart discussed nearby (\ref{eq:def_P}) and (\ref{eq:def_S}).} For convenience we work with complex objects\footnote{We use $Z^{ab}=E^{ab}+\imath B^{ab}$ and $V^{ab}=S^{ab}+\imath P^{ab}$. See appendices \ref{sec:grav_E_B} and \ref{sec:Scalar_spin_2} for more details.}
			\begin{eqnarray}
				h_{G}&=&h_{GE}-h_{GB}=\Re \int_{\Sigma} Z^{a b} (-\triangle^{-1}) V_{ab} \label{eq:def_H_z}\\
				\widetilde{h}_{G}&=&h_{GE}+h_{GB}= \Re \int_{\Sigma} Z^{a b} (-\triangle^{-1}) \overline{V}_{ab} \label{eq:H_Z_bar}
			\end{eqnarray}
			 We list a few properties which support our hypothesis that (\ref{eq:def_H_z}) and (\ref{eq:H_Z_bar})  play a role of a `topological charge':
			\begin{itemize}
				\item Similarities with the electromagnetic case:
				\begin{itemize}
					\item Analogy to the electromagnetic helicity -- the quantity (\ref{eq:def_H_z}) in terms of complex scalar field (\ref{eq:H_Z_in_Psi}) is similar to (\ref{eq:h_Phi_1}).
					\item Analogy to the conservation law -- (\ref{eq:def_H_z}) is conserved in time if (\ref{eq:topo_charge_grav_cond}) is fulfilled. It is analogous to (\ref{eq:topo_charge_condition}).
				\end{itemize}
 				\item (\ref{eq:def_H_z}) is conserved in time for an example of gravitational type N hopfion described in \cite{thompson2015classification}.
 				\item Structure comparable to other quantities defined for linearized gravity field, for example the energy (\ref{eq:Hxy}). 				
			\end{itemize}
		To highlight the analogy with electromagnetic field we express equation (\ref{eq:def_H_z}) in terms of complex scalar field. Using the reduction presented in appendices \ref{sec:spher_ident} and \ref{sec:Scalar_spin_2} the result is as follows:
		\begin{eqnarray}
			h_{G}&=&-\Re \int_{\Sigma}  \imath \Psi \mathbf{\Delta}^{-1} \left(\mathbf{\Delta}+2\right)^{-1} \partial_{t} \Psi \label{eq:H_Z_in_Psi} \\
			\widetilde{h}_{G}&=&\frac{1}{2}\int_{\Sigma} \Re \left[ \imath \left(\Psi \mathbf{\Delta}^{-1}\left(\mathbf{\Delta}+2\right)^{-1} \partial_{t} \overline{\Psi}-\overline{\Psi}\mathbf{\Delta}^{-1}\left(\mathbf{\Delta}+2\right)^{-1} \partial_{t} \Psi \right) \right]
		\end{eqnarray}
		If we compare $\partial_{t} h_{G}$ and the real part of $\int_{\Sigma}Z^{kl} \triangle^{-1} Z_{kl}$ in terms of the scalar $\Psi$ then turns out that they are equal up to the factor $2$
		\begin{equation}
			\partial_{t} h_{G}=-2\Re\int_{\Sigma}\imath Z^{kl} (-\triangle^{-1}) Z_{kl}=-\Re \int_{\Sigma}  \imath \partial_{t}\left( \Psi \mathbf{\Delta}^{-1} \left(\mathbf{\Delta}+2\right)^{-1} \partial_{t} \Psi \right) \label{eq:H_Z_preservance}
		\end{equation}
		 The gravitational helicity $\widetilde{h}_{G}$ is preserved in time for all $\Psi$ which fulfill wave equation
		 \begin{equation}
		  \partial_{t} \widetilde{h}_{G}=\frac{1}{2}\int_{\mathcal{S}}  \Re \left[\imath \left(\Psi\mathbf{\Delta}^{-1} \left(\mathbf{\Delta}+2\right)^{-1} r^2 \partial_{r} \overline{\Psi}-  r^2 \partial_{r} \Psi\mathbf{\Delta}^{-1} \left(\mathbf{\Delta}+2\right)^{-1} \overline{\Psi} \right)\right] \label{eq:tilde_H_Z_preservance}
		 \end{equation}
		 We assume the linearized gravity fields are localized\footnote{By localized we mean compactly supported or with fall off sufficiently fast which enables one to neglect boundary terms.}. The results (\ref{eq:H_Z_preservance}) and (\ref{eq:tilde_H_Z_preservance}) give
		 \begin{sthm}
		 	For localized fields, the objects $h_{GE}$ (\ref{eq:def_H_E}) and $h_{GB}$ (\ref{eq:def_H_B}) are preserved in time if and only if
		 	\begin{equation}
		 	 2 \Re\int_{\Sigma}Z^{kl} (-\triangle^{-1}) Z_{kl}=\Re \int_{\Sigma}  \partial_{t}\left( \Psi \mathbf{\Delta}^{-1} \left(\mathbf{\Delta}+2\right)^{-1} \partial_{t} \Psi \right)=0 \label{eq:topo_charge_grav_cond}
		 	\end{equation}
		 \end{sthm}
The above theorem corresponds to proposition \ref{sthm:E_M_topo_conservation} in electrodynamics.
\section{Conclusions}
In the paper, the electromagnetic hopfions are described in terms of the complex scalar $\Phi$
which contains the full information about Maxwell field --- two unconstrained degrees of freedom. The scalar $\Phi$ formalism for electrodynamics is presented in appendix \ref{Sec:scal_rep_EM_field}. We generalize the electromagnetic hopfions by the natural generalization\footnote{The scalar $\Phi$ for hopfions is equal to $\frac{r Y_{1}}{\left(r^2-(t-\imath)^2\right)^{2}}$. The type of hopfion, namely null or non-null (see \cite{thompson2015classification}) is related to the order of the dipole.} of $\Phi$ to the higher multipole solution (\ref{eq:Phi_Hopf}). The physical quantities, like energy or helicity, are expressed in terms of the scalar.\\
The electromagnetic case can be treated as a ``toy-model'' for the linearized gravity.
Next, the scalar $\Psi$ description of gravito-electromagnetic formulation of linearized gravity is presented (appendix \ref{sec:Scalar_spin_2}). In analogy to electromagnetism, we generalize the gravitational hopfion to higher multipole solution (\ref{eq:Psi_hopf}). We propose the notion of helicity for linearized gravity $h_{GE}$ (\ref{eq:def_H_E}) and $h_{GB}$ (\ref{eq:def_H_E}). The properties of gravitational helicities in terms of the scalar $\Psi$ are similar to electromagnetic ones. We compare gravitational quasi-local densities for quadrupole solution (\ref{eq:Psi_solut_quadru}). The results are presented and discussed in section \ref{sec:energy_comparision}.\\
The structure of the theory for electromagnetism and linearized gravity can be illustrated on the diagram \ref{diag:Comparision theories}:
\captionsetup[table]{name=Diagram}
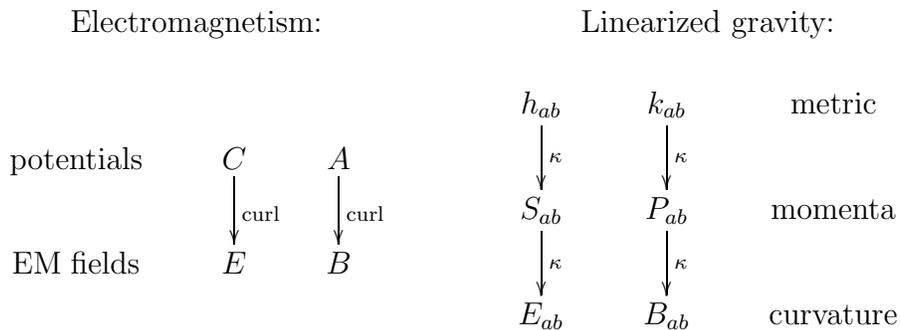
\begin{table}[ht]
	\centering
	\begin{tabular}{cc}			
		Electromagnetism: & Linearized gravity:\\	
		\begin{minipage}[r]{0.45 \textwidth}			
			\begin{equation*}
			\xymatrix{
				{\rm potentials}& C \ar[d]^{\curl} &A\ar[d]^{\curl}\\
				{\rm EM \; fields} &E  &B}
			\end{equation*}
			
		\end{minipage}
		
		&
		\begin{minipage}[l]{0.4 \textwidth}
			\begin{equation*}
			\xymatrix{
				h_{ab} \ar[d]^\kappa  &k_{ab}\ar[d]^\kappa & {\rm metric}\\		
				S_{ab} \ar[d]^\kappa  &P_{ab}\ar[d]^\kappa & {\rm momenta}\\
				E_{ab} &B_{ab} & {\rm curvature}}
			\end{equation*}
		\end{minipage}  \\
	\end{tabular}
	\caption{Comparison of the structures of electromagnetism and linearized gravity. Where $h_{ab}$ and $k_{ab}$ are respectively the linearized metric and its dual companion. $\kappa$ is the first order differential operator. For transverse-traceless gauge, $\kappa$ is simply the symmetric curl operator for symmetric tensors.}
	\label{diag:Comparision theories}
\end{table}

We would like to point out the following:
\begin{enumerate}
	\item Spin-2 field theory (see appendix C.2) starts with linearized Weyl tensor as a primary object and Bianchi identities play a role of evolution equations. Theory of linearized gravity has a richer structure. It contains ``potentials'' for curvature tensors: metric, momenta and their dual counterparts. See rhs of diagram \ref{diag:Comparision theories}. That simple observation has consequences for (non-)locality of densities of energy and helicity. %discussed below.
	
	\item The energy functional for Maxwell theory is constructed from electromagnetic vector fields $E$ and $B$. The energy density is local at a point in terms of $E$ and $B$.
	In the case of linearized gravity
	%the natural objects  in which
	the Hamiltonian energy density (see (\ref{eq:Hxy})) becomes local
	as a combination of the metric and curvature. However, in terms of spin-2 field,
	the energy functional (\ref{eq:Hxy}) contains non-local $Z^{kl} (- \triangle)^{-1} Z_{kl}$ term.
	More precisely, the  object $(- \triangle)^{-1} Z_{kl}=\left(\kappa^{-1}\right)^{2} Z_{kl}$ is locally related to a combination of metrics $h_{ab}$ and $k_{ab}$ (see rhs of diagram \ref{diag:Comparision theories}).
	Another form of the energy functional ($\left(\kappa^{-1} Z_{kl}\right)^{2}$ -- square of momenta) contains non-local integral operator $\kappa^{-1}$ which is responsible for the non-locality of energy density described by (\ref{eq:Hxy}).
	
	\item For helicity of linearized gravity the similar problems occur like for energy. The natural objects for helicity functional to be local are metric and momenta.

\end{enumerate}
The precise description of $\kappa$ operator and the structure of linearized gravity %in the presented framework
will be presented in a separate paper \cite{jjjkenergy}.
	\vspace{0.5 cm}\\
	{\noindent \sc Acknowledgements} This work was supported in part by Narodowe Centrum Nauki (Poland) under Grant No. 2016/21/B/ST1/00940.
	
  \appendix
  	
		\section{Mathematical supplement \label{sec:math_sup}}
			\subsection{Three-dimensional Laplace operator and its inverse}
				Consider Laplace equation
				\begin{equation}
				\triangle G(\mathbf{r},\mathbf{r}')= - \delta^{(3)}(\mathbf{r}-\mathbf{r}') \label{eq:3_D_laplace}
				\end{equation}
				with a solution on an open set without boundary. $\delta^{(3)}(\mathbf{r}-\mathbf{r}')$ is a three-dimensional Dirac delta. $G(\mathbf{r},\mathbf{r}')$ is the following Green function of (\ref{eq:3_D_laplace})
				\begin{equation}
					G(\mathbf{r},\mathbf{r}')=\frac{1}{4 \pi ||\mathbf{r}-\mathbf{r}'||}
				\end{equation}
				The solution of Poisson equation
				\begin{equation}
				\triangle u(\mathbf{r})=-f(\mathbf{r})
				\end{equation}
				is the convolution of $f(\mathbf{r})$ and Green function
				\begin{equation}
				u(\mathbf{r})=\int_{\Sigma} f(\mathbf{r}') G(\mathbf{r},\mathbf{r}') \mathrm{d} \mathbf{r}'=\int_{\Sigma} \frac{f(\mathbf{r}')}{4 \pi ||\mathbf{r}-\mathbf{r}'||} \mathrm{d} \mathbf{r}'
				\end{equation}
				\subsection{Two-dimensional Laplace operator and its inverse}
				Consider two-dimensional unit sphere in $\mathbb{R}^3$, parameterized by  a unit position vector $n$. One of the main differences is that the domain of the solutions is the compact surface without boundary. The conclusions of the Stokes theorem $(\int_{\mathbb{S}^2} \mathbf{\Delta} u(n)=0)$ require a modified problem to be examined than in the three-dimensional case. Consider the following two-dimensional Laplace equation with an additional condition
				\begin{eqnarray}
						\mathbf{\Delta} \mathbf{G}(n,n')&=&1- \delta^{(2)}(n-n') \label{eq:2_D_laplace} \\
						\int_{\mathbb{S}^2} \sigma \mathbf{G}(n,n') \mathrm{d} n'&=&0
				\end{eqnarray}
				where $\sigma$ is area element on $\mathbb{S}^2$. We have the solution
				\begin{equation}
				\mathbf{G}(n,n')=-\frac{1}{4 \pi}\left(\ln(\frac{1-n \cdot n'}{2})+1\right)
				\end{equation}
				where `$\cdot$' is a scalar product of the position vectors\footnote{For a given point $(\theta, \varphi)$ in spherical coordinates on the unit sphere, the three-dimensional position vector in the Cartesian embedding is $n=\sin \theta \cos \varphi \partial_{x}+ \sin \theta \sin \varphi \partial_{y}+\cos \theta \partial_{z}$. Then we use scalar product with Euclidean metric.}. The solution of the Poisson equation
				\begin{eqnarray}
					\mathbf{\Delta} s(n)&=&-f(n)  \\
					\int_{\mathbb{S}^2} \sigma f(n) \mathrm{d} n&=&0
				\end{eqnarray}
				is the convolution of $f(n')$ and the Green function
				\begin{equation}
						u(n)=\int_{\mathbb{S}^{2}} f(n') \mathbf{G}(n,n') \mathrm{d} n'=-\int_{\mathbb{S}^{2}}\frac{1}{4 \pi}\left(\ln(\frac{1-n \cdot n'}{2})+1\right)  f(n') \mathrm{d} n'
				\end{equation}
				 See \cite{CJM1998Uni} and \cite{jezierski2002peeling} for detailed view, \cite{szmytkowski2006closed} is a specialized literature on the subject.
			\subsection{Operations on the sphere \label{sec:operation_on_S2}}
			 Let us denote by $\mathbf{\Delta}$ the Laplace--Beltrami operator associated with the standard metric $h_{AB}$ on $S^2$. Let $SH^l$ denote the space of
			spherical harmonics of degree $l$ ($g\in SH^l \Longleftrightarrow
			{\mathbf{\Delta}}g= -l(l+1)g$).
			Consider the following sequence
			\[
			\begin{array}{ccccccccc}
			V^0\oplus V^0 & \stackrel{i_{01}}{\longrightarrow} & V^1
			& \stackrel{i_{12}}{\longrightarrow} & V^2
			& \stackrel{i_{21}}{\longrightarrow} & V^1
			& \stackrel{i_{10}}{\longrightarrow} & V^0 \oplus V^0 %\ .
			\end{array}
			\]
			Here $V^0$ is the space of, say, smooth functions on $S^2$,
			$V^1$ -- that of smooth covectors on $S^2$, and
			$V^2$ -- that of symmetric traceless tensors on $S^2$.
			The various mappings above are defined as follows:
			\[ i_{01}(f,g)=f_{||a}+\varepsilon_a{^b}g_{||b} \]
			\[ i_{12}(v)= v_{a||b}+ v_{b||a}-h_{ab}v^c_{||c} \]
			\[ i_{21}(\chi)= \chi_a{^b}{_{||b}} \]
			\[ i_{10}(v)=\left( {v^a}_{||a}, \varepsilon^{ab}v_{a||b} \right) \]
			where $||$ is used to denote the covariant derivative with respect to
			the Levi--Civita connection of the standard metric $h_{AB}$ on
			$S^2$. For more details see appendix E in \cite{jezierski2002peeling}.
			\subsection{Identities on the sphere \label{sec:spher_ident}}
			 We have used the following identities on a sphere
			\begin{equation}
				-\int_{S(r)} \pi^A v_A = \int_{S(r)} (r{\pi}^{A}{_{|| A}}) {\mathbf{\Delta}} ^{-1}
				(rv^{A}{_{|| A}})+ \int_{S(r)} (r{\pi}^{A|| B}\varepsilon_{AB}) {\mathbf{\Delta}} ^{-1}
				(rv_{A|| B}\varepsilon^{AB})
			\end{equation}
			and similarly for the traceless tensors we have
			\begin{eqnarray}
				\int_{S(r)} \stackrel{\circ}{\pi}\! ^{AB} \stackrel{\circ}{v}\! _{AB} &=&
				2 \int_{S(r)} (r^2\varepsilon^{AC}
				\stackrel{\circ}{\pi}\! {_A{^B}}{_{||BC}})
				{\mathbf{\Delta}} ^{-1}({\mathbf{\Delta}}+2) ^{-1}
				(r^2\varepsilon^{AC}\stackrel{\circ}{v}\! {_A{^B}}{_{|| BC}})  \nonumber \\
				& & +\, 2 \int_{S(r)} (r^2\stackrel{\circ}{\pi}\! {^{AB}}{_{|| AB}})
				{\mathbf{\Delta}} ^{-1}({\mathbf{\Delta}}+2) ^{-1}
				(r^2\stackrel{\circ}{v}\! {^{AB}}{_{|| AB}})
			\end{eqnarray}
			
		\section{Scalar representation of electromagnetic field \label{Sec:scal_rep_EM_field}}
			Let us consider an electromagnetic field on Minkowski background. We present how to describe electromagnetism in terms of complex scalar function $\Phi$. The section is organized as follows: we start with a description of  standard electric $E$ and magnetic $B$ fields with  help of complex Riemann--Silberstein vector $Z=E+\imath B$. Then, we decompose $Z$, in the spherical coordinate system, into radial and angular part. We show that the radial part is sufficient to recover quasi-locally the whole $Z$ vector.\\
			 The vacuum Maxwell equations for electric field vector $E$, magnetic field $B$, and vector potential $A$ are
  			\begin{eqnarray}
  			\ddiv E&=&0 \label{eq:Maxwell_1} \\
  			\ddiv B&=&0 \\
  			\curl E&=&- \frac{\partial B}{\partial t}\\
  			\curl B&=& \frac{\partial E}{\partial t}\\
  			B&=&\curl A \label{eq:Maxwell_5}
  			\end{eqnarray}
  			If electric field $E$ is sourceless then additional vector potential can be introduced\footnote{We remark that the description of electromagnetism with the help of complex scalar function holds also without the additional potential.}. It is defined up to a gradient of a function in the following way
  			\begin{equation}
  			E=\curl C \label{eq:def_C}
  			\end{equation}
  			It is convenient to use one complex electromagnetic vector field $Z$, called Riemann--Silberstein vector, instead of $E$ and $B$. $Z$ is defined as follows
  			\begin{equation}
  			Z=E+\imath B \label{eq:RS}
  			\end{equation}
  			where $\imath^2=-1$. For sake of simplicity, we will use complex vector potential $V$ instead of $C$ and $A$:
  			\begin{equation}
  			V=C+\imath A \label{eq:def_W}
  			\end{equation}
  			The vacuum Maxwell equations (\ref{eq:Maxwell_1}--\ref{eq:Maxwell_5}) with vector potential $C$ (\ref{eq:def_C}) can be written in the form of three complex, differential equations for vector fields
  			\begin{eqnarray}
  			\ddiv Z&=&0 \label{eq:cMaxwell1} \\
  			\curl Z&=&\imath \frac{\partial Z}{\partial t} \label{eq:cMaxwell2}\\
  			Z&=& \curl V \label{eq:cMaxwell3}
  			\end{eqnarray}
  			In the next part of the section, we will use spherical coordinate system. Each vector $w=(w^{r},w^{A})$ can be decomposed into its radial part $w^{r}$ and two-dimensional angular part $w^{A}$. The capital letter index runs angular coordinates.\\
  			We split two-dimensional vector into its longitudinal ${w^{A}}_{||A}$  and transversal part  $\varepsilon_{r AB}{w^{A||B}}$. The Maxwell equations in terms of the decomposition have the form:
  			\begin{eqnarray}
  				r Z^{r}&=&\Phi \\
  				r^{2}Z^{A}_{||A}&=&- \partial_{r}(r \Phi) \\
  				r \varepsilon_{r AB}Z^{A||B}&=&- \imath \partial_{t} \Phi \\
  				\mathbf{\Delta} V_{r}-{V_{C,r}}^{||C}&=&-\imath \partial_{t} \Phi\\
  				r \varepsilon_{r AB}V^{A||B}&=&-\Phi
  			\end{eqnarray}
  			\subsection{Scalar representation of electromagnetic field in curved spacetimes}
  			%\cmti{New}
  			The description of electromagnetism in terms of complex scalar field can be generalized for type D spacetimes (Petrov classification). For example, in \cite{jezsmo} it has been done for Kerr spacetime. The generalization of $\Phi$ %(\ref{eq:def_Phi})
  for Kerr, we denote it by $\Phi_{K}$, is constructed from conformal Yano--Kiling tensor $Q_{\mu \nu}$, its dual companion $\ast Q_{\mu \nu}$ and Maxwell field $F_{\mu \nu}$. $\ast$ denotes the Hodge duality. The contraction
  \[ \Phi_{K}:=\frac{\imath}{2} F^{\mu \nu} \left[Q_{\mu \nu}- \imath(\ast Q_{\mu \nu})\right] \]
   fulfills extended wave equation
  				\begin{equation}
  			\widetilde{\Box} \Phi_{K} + \frac{2 m}{(r- \imath a \cos \theta)^3} \Phi_{K}=0
  			\end{equation}
  			which is called Fackerell--Ipser equation. $\widetilde{\Box}$ is d'Alembert operator for Kerr metric. $m$ and $m a$ are respectively mass and angular momentum of Kerr black hole. $r$ and $\theta$ belong to Boyer--Lindquist coordinates. For detailed results and discussion see \cite{jezsmo}.
  			\subsection{Simple observation}
  			Let us consider four-dimensional Laplace equation in the four-dimensional Euclidean space
  			\begin{equation}
  			\stackrel{(4)}{\triangle} f(x)=-\delta(x) \label{eq:4D_Laplace_eq}
  			\end{equation}
  			where $\delta(r)$ is the Dirac delta. We focus on the following solution of (\ref{eq:4D_Laplace_eq}) given  in the Cartesian coordinates: %it takes the form
  			\begin{equation}
  			f(x)=\frac{1}{(x^{0})^{2}+||\mathbf{r}||^{2}} \label{eq:4_D_solution}
  			\end{equation}
  			where $||\mathbf{r}||=r=\sqrt{\sum_{i=1}^{3} x_{i}^{2}}$. The solution (\ref{eq:4_D_solution}) can be extended analytically on Minkowski spacetime by the transformation
  			\begin{equation}
  				x^{0}=\imath t +1
  			\end{equation}
  			We receive
  			\begin{equation}
  			\widetilde{f}(t,r)=\frac{1}{r^2-(t-i)^2}
  			\end{equation}
  			which fulfills wave equation on Minkowski background. Now, consider  $l$-th order differential operator $A_{l}$ which
  			\begin{enumerate}
  				\item generates a function $F_{l}(t,r)$ from $\widetilde{f}(t,r)$ which is proportional to $l$-th spherical mode\footnote{$l$-th mode from spherical harmonics decomposition which fulfills $\mathbf{\triangle} F_{l}=-l(l+1) F_{l}$},
  				\item commutes with d'Alembert operator.
  			\end{enumerate}
  		For $l=1$ we have the following example
  		\begin{equation}
  		A_{1}=\frac{\partial}{\partial x_{1}} \pm  \imath \frac{\partial}{\partial x_{2}}
  		\end{equation}
  		and
  		\begin{eqnarray}
  		0&=&A_{1} \Box \widetilde{f}(t,r)  \nonumber \\
  		&=&\Box A_{1} \widetilde{f}(t,r) \nonumber \\
  		&=& \Box F_{1}(t,r)
  		\end{eqnarray}
  		hence $F_{1}(t,x)$ fulfills wave equation. The above simple observation enables one to generate a solution similar (up to a constant) to (\ref{eq:Phi_Hopf}).

	\section{Scalar description of linearized gravity \label{sec:section_lin_grav}}
		\subsection{Equivalent definitions of spin-2 field}
		Let us start with the standard formulation of a spin-2 field $W_{\mu \alpha
			\nu \beta}$ in the Minkowski spacetime equipped with a flat
		metric $g_{\mu\nu}$ and its inverse $g^{\mu\nu}$. We consider vacuum case.
		The field $W$ can be also
		interpreted as a Weyl tensor for linearized
		gravity (see \cite{christodoulou1990asymptotic}, \cite{jezierski1995relation}, \cite{jezierski2002peeling}).

		The following algebraic properties:
		\begin{equation}
		\label{s2W} W_{\mu \alpha \nu \beta}=W_{\nu \beta \mu \alpha} =
		W_{[\mu \alpha ] [ \nu \beta ]}
		\; , \; \;  W_{\mu [\alpha \nu \beta ]}=0 \; , \; \;
		g^{\mu\nu}W{_{\mu\alpha \nu \beta}} =0
		\end{equation}
		and Bianchi identities which play a role of field equations
		\begin{equation}
		 \stackrel{(4)}{\nabla}_{[\lambda} W_{\mu\nu ] \alpha\beta} =0
		\end{equation}
		can be used as a definition of spin-2 field $W$.				
		The $*$--operation defined as
		\[ ({^*} W)_{\alpha\beta\gamma\delta}=\frac 12 \varepsilon_{\alpha \beta
			\mu \nu} W^{\mu\nu}{_{\gamma\delta}} \, , \quad
		(W^*)_{\alpha\beta\gamma\delta}=\frac 12 W_{\alpha\beta}{^{\mu\nu}}
		\varepsilon_{\mu \nu \gamma\delta}  \]
		has the following properties:
		\[ ({^*} W^*)_{\alpha\beta\gamma\delta}=\frac 14 \varepsilon_{\alpha \beta
			\mu \nu} W^{\mu\nu\rho\sigma}\varepsilon_{\rho\sigma\gamma\delta} \, , \quad
		{^*} W =W^* \; , \; \; {^*} ({^*} W) = {^*}W^* =-W \]
		where $\varepsilon_{\mu \nu \gamma\delta}$ is a Levi--Civita
		skew-symmetric tensor\footnote{Defined in footnote \ref{L-C_tensor}.}
		and ${^*}W$ is called dual spin-2 field.
		The above formulae are also valid for general Lorentzian metrics.
		\subsection{Gravito-electric and gravito-magnetic formulation \label{sec:grav_E_B}}
		Following Maartens \cite{maartens1998gravito}, spin-2 field can be equivalently described in terms of gravito-electric and gravito-magnetic tensors. We perform a $(3+1)$--decomposition of the Weyl tensor. The ten independent components of $W$ split into two three-dimensional symmetric, traceless tensors: the electric part
		\begin{equation}
		E(X,Y):=W(X,\partial_{t},\partial_{t},Y)
		\end{equation}
		and the magnetic part
		\begin{equation}
		B(X,Y):={^*}W(X,\partial_{t},\partial_{t},Y)
		\end{equation}
		The following relations between $W$ and the three-dimensional tensors hold:
		\begin{eqnarray}
		W_{0kl0}=E_{kl}\, , & W_{0kij}=B_{kl}\varepsilon^l{_{ij}} \, , &
		W_{klmn}= \varepsilon^i{_{kl}}\varepsilon^j{_{mn}}E_{ij} %\, .
		\end{eqnarray}
		The classical formulation of gravito-electromagnetism uses the constraint equations
					\begin{eqnarray}
		{E^{kl}}_{|l}&=&0 \label{eq:constrain_E}\\
		{B^{kl}}_{|l}&=&0
		\end{eqnarray}
		and the dynamical equations
		\begin{eqnarray}
			\partial_{t} E^{kl}&=&  \varepsilon^{pq(k}{B^{l)}}_{q|p} \\
			\partial_{t} B^{kl}&=& -\varepsilon^{pq(k}{E^{l)}}_{q|p}
		\end{eqnarray}
		where $\left[\curl X \right]_{ab}:=\varepsilon_{cd(a}{X_{b)}}^{d|c}$ is the symmetric curl operator for tensors.
		 \paragraph{ADM momentum $P$ and the dual counterpart $S$ as ``potentials'' for Weyl tensor.} Analogically to electromagnetic case we introduce potentials for Weyl tensor in gravito-electromagnetic formulation. The potential for gravito-magnetic part is the ADM momentum $P$. It fulfills
		\begin{equation}
		B_{ab}= \varepsilon_{cd(a}{P_{b)}}^{d|c} \label{eq:def_P}
		\end{equation}
		The second potential can be introduced for gravito-electrical part
		\begin{equation}
		E_{ab}=\varepsilon_{cd(a}{S_{b)}}^{d|c} \label{eq:def_S}
		\end{equation}
		The potentials fulfill constraint equations
			\begin{eqnarray}
			 {P^{kl}}_{|l}&=&0 \\
			 {S^{kl}}_{|l}&=&0 \label{eq:constrain_S}
			 \end{eqnarray}
		It is convenient to use a complex combination of $E_{kl}$ and $B_{kl}$ as follows
		\begin{equation}
		Z_{kl}:=E_{kl}+\imath B_{kl}
		\end{equation}
		and its potentials $P_{kl}$ and $S_{kl}$
		\begin{equation}
		V_{kl}=S_{kl}+\imath P_{kl}
		\end{equation}
		The equations (\ref{eq:constrain_E})-(\ref{eq:constrain_S}) in terms of complex objects are
		\begin{eqnarray}
			{Z^{kl}}_{|l}&=&0 \label{eq:Z_costrain} \\
			\dot{Z}^{kl}&=& -\imath \varepsilon^{pq(k}{Z^{l)}}_{q|p} \label{eq:Z_dynamical} \\
			Z_{ab}&=&\varepsilon_{cd(a}{V_{b)}}^{d|c} \label{eq:V_grav_curl} \\
			{V^{kl}}_{|l}&=&0 \label{eq:V_grav_div}
		\end{eqnarray}
		\subsection{Scalar representation of spin-2 field \label{sec:Scalar_spin_2}}
		Spin-2 field can be represented as a complex, scalar function defined analogically to the electromagnetic case\footnote{See the equation (\ref{eq:wave_eq}) and the comments below.}. In the spherical coordinates it has the form
		\begin{equation}
		\Psi=2 Z_{kl}x^k x^l=2 Z_{rr}r^2
		\end{equation}
		The recovery procedure of the $Z_{kl}$ field from $\Phi$ uses the  constraint equations for linearized Weyl tensor and the dynamical equations.
		The $(2+1)$--splitting of the constraint (\ref{eq:Z_costrain}):
		\begin{eqnarray}
			\partial_r(r^3Z^{rr}) + r^3 Z^{rA}{_{||A}}&=&0 \label{eq:constrain_Z^{rr}}\\
			\partial_r(r^4Z^{rA}{_{||A}}) + r^4\kolo{Z}^{AB}{_{||AB}} -\frac12 r^2\mathbf{\Delta} Z^{rr} &=& 0\\ \label{eq:constrain_Z_polar}
			\partial_r(r^4 Z^{r}{_{A||B}}\varepsilon^{AB})+ r^4\kolo{Z}_A{^B}{_{||BC}}   \varepsilon^{AC} &=&0
		\end{eqnarray}
		and the $(2+1)$--decomposition of the dynamical equations (\ref{eq:Z_dynamical}) :
		\begin{eqnarray}
		\partial_{t} Z^{rr}&=&-r^2 \varepsilon^{rAB}Z_{r A ||B} \\
		\partial_{r}(r^2 \partial_{t} Z^{rr})&=&- \imath r^{4} {{\kolo{Z}_{A}}^{B}}_{||B C} \varepsilon^{r A C}
		\end{eqnarray}		
		enables one to express explicitly all electromagnetic components of the Weyl tensor in terms of $\Psi$ and $\partial_{t} \Psi$:
		\begin{eqnarray}
		r^2 Z^{rr} &=& \frac{1}{2} \Psi \label{eq:2D_Z_eqs_1} \\
		r^2 Z_{rA||B}\varepsilon^{rAB} &=& -\frac{1}{2} \imath\partial_{t} \Psi\\
		r^3 Z^{rA}{_{||A}}&=&-\frac{1}{2} \partial_r(r{ \Psi}) \\
		r^2 \stackrel{(2)}{Z}&=&-\frac{1}{2} \Psi \\
		r^4 \stackrel{\circ}{Z}^{AB}{_{||AB}} &=& \frac{1}{2} \partial_r\left(r\partial_r(r{\Psi}) \right) + \frac14 \mathbf{\Delta} \Psi \\
			r^4 \stackrel{\circ}{Z}_A{^B}{_{||BC}}   \varepsilon^{rAC} &=& \frac{1}{2} \imath \partial_r (r^2 \partial_{t} \Psi) \label{eq:2D_Z_eqs_2}
		\end{eqnarray}
		where $\stackrel{(2)}{Z}=g^{AB}Z^{AB}$ and $\stackrel{\circ}{Z}_{AB}=Z_{AB}-g_{AB}\stackrel{(2)}{Z}$.
		The scalar is related to a gauge-independent part of the potential $V_{ab}$.
		The $(2+1)$ -- splitting of (\ref{eq:V_grav_curl}), (\ref{eq:V_grav_div}) and use of (\ref{eq:2D_Z_eqs_1}--\ref{eq:2D_Z_eqs_2}) gives
		\begin{eqnarray}
			\mathbf{\Delta}(\mathbf{\Delta}+2){V^{r}}_{r}&=&-\left(2 \imath \partial_{t} \Psi+2 \imath \left(r \Pi\right)_{,r}+ \imath (\mathbf{\Delta}+2) \Pi\right) \\
			(\mathbf{\Delta}+2){V^{rA}}_{||A}&=&\imath \frac{\partial_{t} \Psi +(r \Pi)_{,r}}{r} \\
			2 r^{2} V^{r A||B}\varepsilon_{r A B}&=&- \Psi\\
			r^{2} \vphantom{V}\stackrel{{(2)}}{V}\!\vphantom{V}&=& \Psi\\
			2 r^2 {\kolo{V}^{AB}}_{||AB}&=&-\imath\left(\partial_{t} \Psi -\Pi\right)\\
			2 r^{4} {\kolo{V}^{C}}_{A||CB} \varepsilon^{r A B}&=&\left(r^{2} { \Psi}\right)_{,r}		
		\end{eqnarray}
		where $\Pi=2 r {V^{rA}}_{||A}+\mathbf{\Delta} {V^{r}}_{r}$ is a gauge dependent part.
	\bibliography{BibTex_hopfion}
\end{document}